\documentclass[12pt]{article}
\usepackage{epsf,amsfonts,amssymb,amsbsy}
\def\slashchar#1{\setbox0=\hbox{$#1$}           % set a box for #1
   \dimen0=\wd0                                 % and get its size
   \setbox1=\hbox{/} \dimen1=\wd1               % get size of /
   \ifdim\dimen0>\dimen1                        % #1 is bigger
      \rlap{\hbox to \dimen0{\hfil/\hfil}}      % so center / in box
      #1                                        % and print #1
   \else                                        % / is bigger
      \rlap{\hbox to \dimen1{\hfil$#1$\hfil}}   % so center #1
      /                                         % and print /
   \fi}                                         %
\begin{document}
\makeatletter
\def\section{\@startsection {section}{1}{\z@}{-3.5ex plus-1ex minus
    -.2ex}{2.3ex plus.2ex}{\reset@font\small\bf\uppercase}}
\makeatother

%\vspace*{4cm}
\title{\large \bf Electroweak symmetry breaking in Higgs mechanism with
composite operators and solution of naturalness }
\author{V.V. Kiselev,\\
{\small\it State Research Center of Russia "Institute for High
Energy Physics"},\\ {\small\sl Protvino, Moscow region, 142281 Russia}\\
{\small\sl E-mail: kiselev@th1.ihep.su, Fax: +7-(0965)-744739 }}
\date{}

%\begin{center}
%{\large \bf Electroweak symmetry breaking in Higgs mechanism with composite
%operators and solution of naturalness
%}\\
%\vspace*{5mm}
%V.V. Kiselev
%\end{center}
%\begin{center}
%{\it State Research Center of Russia
%"Institute for High Energy Physics",\\ Protvino, Moscow region,
%142284 Russia\\
%E-mail: kiselev@th1.ihep.su, Fax: +7-0952302337}\\
%\end{center}
\maketitle

\begin{abstract}
{Introducing a source for a bi-local composite operator motivated by the
perturbative expansion in gauge couplings, we calculate its effective potential
in the renormalization group of Standard Model with no involvement of
technicolor. The potential indicates the breaking of electroweak symmetry below
a scale $M$ due to a nonzero vacuum expectation value of neutral component for
the ${\sf SU(2)}$-doublet operator. At virtualities below a cut off $\Lambda$
we introduce the local higgs approximation for the effective fields of sources
coupled to the composite operators. The value of $\Lambda\approx 600$ GeV is
fixed by the measured masses of gauge vector bosons. The exploration of
equations for infrared fixed points of calculated Yukawa constants allows us to
evaluate the masses of heaviest fermion generation with a good accuracy, so
that $m_t(m_t) = 165\pm 4$ GeV, $m_b(m_b) = 4.18\pm 0.38$ GeV and
$m_\tau(m_\tau) = 1.78\pm 0.27$ GeV. After a finite renormalization of
effective fields for the sources of composite operators, the parameters of
effective Higgs field potential are calculated at the scale of matching with
the local theory $\Lambda$. The fixed point for the Yukawa constant of $t$
quark and the matching condition for the null effective potential at $M$ drive
the $M$ value to the GUT scale. The equation for the infrared fixed point of
quartic self-action allows us to get estimates for two almost degenerate scalar
particles with $m_H= 306\pm 5$ GeV,  while third scalar coupled with the $\tau$
lepton is more heavy: $m_{H_\tau} = 552\pm 9$ GeV. Some phenomenological
implications of the offered approach describing the effective scalar field, and
a problem on three fermion generations are discussed.
}
\end{abstract}

\vspace*{1cm}
%\newpage

\section{Introduction}
At present, the Standard Model exhibits almost a total success in experimental
measurements \cite{SM}. The only question being a white spot on its body, is
the empirical verification of mechanism for the spontaneous breaking of
electroweak symmetry. In this respect,  the minimal model involving a single
local Higgs field brings a disadvantage: the stability of potential under the
quantum loop corrections requires a restriction of quadratic divergency in the
self-action by the introduction of ``low'' energy cut-off $\Lambda\sim 10^3$
GeV, which is not a natural physical scale standing far away from what can be
desirable \cite{tHoft}: the GUT scale, $M_{\rm GUT}\sim 10^{16}$ GeV
\cite{GUT}, or even the Planck mass, $M_{\rm Pl}\sim 10^{19}$ GeV. The reason
for putting the $\Lambda$ so small, has to originate beyond the Standard
Model. Two highways to a ``new physics'' merit the most popularity. The first
one is a technicolor \cite{tc} postulating an extra-strong interaction for new
technifermions, which form some ``QCD-like'' condensates, breaking down the
electroweak symmetry and giving the masses to the ordinary gauge bosons.
Despite some problems with the generation of realistic mass values for the
quarks and leptons and suppression of flavor changing neutral currents, the
extended technicolor \cite{etc} provides quite a clear picture for what happens
in the region deeper than $10^3$ GeV. However, the most strict objection
against such the way is the comparison with the current measurements, which
disfavor the technicolor models possessing the calculability \cite{PEP}.
A general consideration of models with the condensation of heavy fermions is
reviewed in ref.\cite{cvet}, while a brilliant presentation of both the
ideas on the electroweak symmetry breaking with composite operators and
techniques as well as results is given in a comprehensive survey by C.T.Hill
and E.H.Simmons \cite{HiSi}. However, the condensation, in general, does not
provide us with the solution of naturalness. In fact, this approach
reformulates the problem as a fine-tuning phenomenon, since the separation of
dynamics responsible for the composite operators at a high scale from the
low-energy electroweak physics takes place at effective couplings tuned to some
critical values. Therefore, we need an additional argumentation in order to
address the naturalness in the framework of condensation mechanism with
composite operators. We present an idea toward this direction below.

The second way is a supersymmetry \cite{SUSY} reforming the
quadratic divergency in the self-action of Higgs field into the
logarithmic one, so that it prescribes the scale $\Lambda$ to be a
splitting between the particles of Standard Model and their
super-partners. Therefore, the supersymmetry has to be broken in a
manner conserving the logarithmic behavior of renormalization,
which is an additional challenge to study and a degree of
ambiguity. However, the advantage is the stability of Higgs
potential, so that $\Lambda$ certainly is a reasonable scale
reflecting the physics in the supersymmetric theory. What remains
is the question: why the basic SUSY scale is so ``low'' in
comparison with the GUT scale? Hence, the naturalness is again the
problem standing in the higher-quality context.

\begin{figure}[th]
\setlength{\unitlength}{1mm}
\begin{center}
\begin{picture}(80,80)
\put(0,0){\epsfxsize=9cm \epsfbox{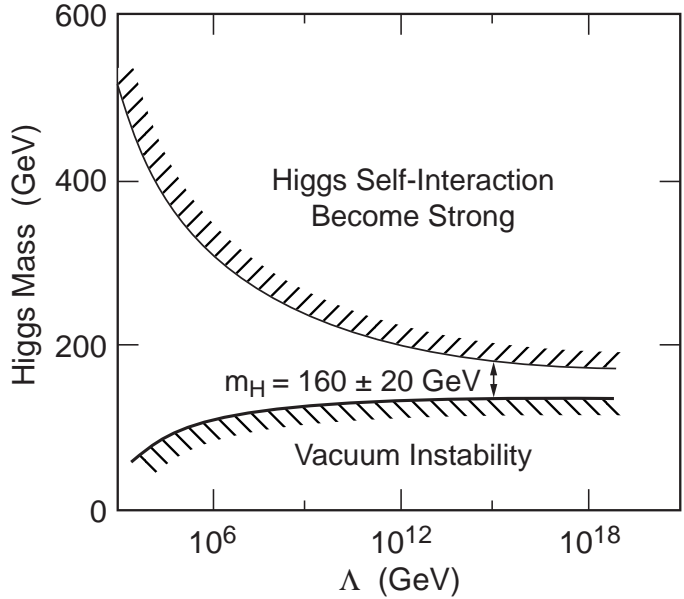}}
\end{picture}
\end{center}
\caption{The region of higgs mass constrained by requirements of the SM
consistency.}
\label{SM}
\end{figure}

If the ultraviolet cut off energy in the loop calculations is placed close to
the Planck scale (see Fig.\ref{SM}\footnote{The figure originally appeared in
ref. \cite{HR}, and it is taken from ref. \cite{Bj}, while the two-loop
consideration recently was done in ref. \cite{PZ}.}), the
Standard Model suffers from the inherent inconsistency except a narrow window
in the range of possible values of Higgs particle mass: $m_H = 160\pm 20$ GeV,
which does not contradict the value following from the precise measurements of
electroweak parameters in the electron-positron annihilation at the $Z$ boson
peak.

The reasons for such the inconsistency are the following: At lower masses the
vacuum stability is broken, i.e., the quartic coupling constant of scalar field
changes its sign \cite{AI}. At higher masses the theory enters the strong
self-interaction regime, which indicates that the quartic coupling constant
becomes infinite (alike the Landau pole) at a scale less than the offered cut
off \cite{Elias}. If the scale of ultraviolet cut off in the SM is much lower
than the Planck scale, then the region of higgs masses providing the SM
consistency, is more wide. However, such the scales are not natural. A low cut
off scale should indicate a new dynamics. While the vacuum instability is an
unavoidable physical constraint, the phase of strong higgs self-interaction
could be treated in the framework of the following representation: The scalar
higgs can be described in terms of the local field below an ultraviolet cut off
$\Lambda$ placed close to the region of strong regime. At virtualities higher
than $\Lambda$, the strongly self-coupled higgs is not fundamental and local
quantum. The dynamics should be described by means of weakly interacting
particles, so that a composite operator with appropriate quantum numbers has to
correspond to the higgs in the `dual' limit implying that the effective
potential of the composite operator yields a development of vacuum expectation
value for the global (independent of space-time point) source of operator. This
strong self-interaction regime could be realized with no involvement of an
extended underlying theory alike the technicolor dynamics, since some composite
operators can develop an appropriate effective potential in the framework of
standard electroweak symmetry.

Our assumptions are the followings:

1. We choose a form of composite operators describing the nonlocal phase of
higgs in the strong self-interaction regime (SSIR) and suppose the connection
of such the operators to the higgses. The suggestion on the nonlocality of
higgses allows us to replace the strong self-interaction regime in the theory
with the local Higgs fields by the weak self-interaction regime (WSIR) of
sources for the composite operators.

2. The interactions of fermions and gauge bosons in the SSIR are given by the
dynamics of SM with no local scalar higgses as well as no extensions like a
technicolor or so.

3. Concerning the position of scale $\Lambda$ denoting the infrared cut off in
the calculations with the composite operators as well as the ultraviolet cut
off in the local theory with the scalar higgs, we put it into the (infrared)
fixed point for the Yukawa coupling constants of heaviest fermion in the local
theory. the numerical value of $\Lambda$ is given by the masses of
weak-interaction gauge bosons.

4. We consider Yukawa couplings of the only heaviest fermion generation in the
SM.

5. In the SSIR we introduce the ultraviolet cut off $M\gg \Lambda$. At $M$ the
electroweak symmetry is exactly restored.

6. We match the effective potential of sources for the composite operators with
the potential of corresponding local scalar fields at the scale $\Lambda$.

\begin{figure}[th]
\setlength{\unitlength}{1mm}
\begin{center}
\begin{picture}(100,45)
\put(0,0){\epsfxsize=11cm \epsfbox{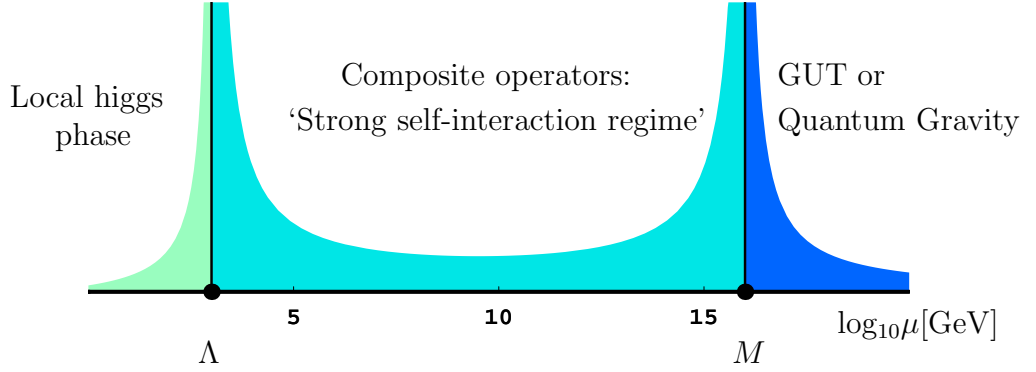}}
\put(15,-4){$\Lambda$}
\put(86,-4){$M$}
\put(100,0){log$_{\rm 10}\mu [{\rm GeV}]$}
\put(-10,30){Local higgs}
\put(-4,25){phase}
\put(27,27){`Strong self-interaction regime'}
\put(34,33){Composite operators:}
\put(92,33){GUT or}
\put(92,27){Quantum Gravity}
\end{picture}
\end{center}
\caption{The division of virtualities as accepted in the calculations
throughout of this paper.}
\label{division}
\end{figure}

The corresponding divisions of virtualities are presented in Fig.
\ref{division}.

It is important to stress that the global sources of composite operators
develop the Higgs-like potential in the region of $[\Lambda;M]$, so that the
corresponding couplings of self-interaction as well as Yukawa constants fall
off to zero under the increase of virtuality from $\Lambda$ to $M$. Therefore,
the dynamics of local interactions is perturbative in the region of $[\Lambda;
M]$, while the notion on the ``strong self-interaction regime'', strictly
speaking, concerns for a theory with the local Higgs field, i.e. we replace
$\left.{\rm SSIR}\right|_{\rm local}\to \left.{\rm WSIR}\right|_{\rm
composite}$.

Postponing a supersymmetric extension in a time, in this paper we develop a
new insight into the breaking of electroweak symmetry by means of exploring the
dynamics of SM to calculate an effective potential for a source of bi-local
operator with no technicolor interactions. The physical reasoning for the
choice of operator under study was hinted in ref.\cite{nz}. So, in the second
order of perturbation theory we write down the following contribution to the
action:
\begin{equation}
 i S_{2m} = - \int dx dy\;\;  {\bf T}[\bar L_L(x) \slashchar{B}(x) L_L(x) \cdot
\bar L_R(y) \slashchar{B}(y) L_R(y)]\cdot 4\pi \alpha_Y \cdot
\frac{Y_L}{2}\cdot  \frac{Y_R}{2},
  \label{eq:2}
\end{equation}
where we have introduced the notations $L_L$ for the left-handed doublets and
$L_R$ for the right-handed singlets, $B$ is the gauge field of weak hypercharge
$Y$, $\alpha_Y$ is its coupling constant. Note, that the gauge field of local
${\sf U(1)}$-group is the only one interacting with both the left-handed and
right-handed fermions. If we suggest a nontrivial vacuum correlators with the
characteristic distance $r\sim 1/v$
\begin{eqnarray}
  \langle 0|{\bf T}[ \slashchar{B}(x) L_L(x)\cdot
\bar L_R(y) \slashchar{B}(y)] |0\rangle  &\Rightarrow&
\frac{\delta(x-y)}{v^4} \langle 0| {\bf T}[\slashchar{B}(x) L_L(x)\cdot
\bar L_R(x) \slashchar{B}(x)] |0\rangle \nonumber \\ & \sim &
\delta(x-y)\; v ,
  \label{eq:3}
\end{eqnarray}
supposing that the scales of expectations for $BB$ and $L_L \bar L_R$ are
driven by $v^2$ and $v^3$, respectively, then the Dirac masses of fermions are
determined by the action\footnote{By the way, Eq.(\ref{eq:3a}) implies that
in the SM the neutrino is massless since its right-handed component is
decoupled, $Y_R=0$.}
\begin{equation}
  \label{eq:3a}
  S_{fm} \sim \int dx\; \bar L_L(x) L_R(x)\cdot v\cdot
4\pi \alpha_Y \cdot
\frac{Y_L}{2}\cdot  \frac{Y_R}{2} +{\rm h.c.}
\end{equation}
In this way we extend the SM action by the initial bi-local bare $J$-term
$$
S_{ib}=\int dx dy\; N_J\cdot J(x,y)\; [\bar L_R(x)\;
\underbrace{\slashchar{B}^{\perp}(x) \slashchar{B}^{\perp}(y)} \;
L_L(y)] - \int dx \phi(x) J(x,x)+{\rm h.c.},
$$
where $N_J=\pi\alpha_Y\cdot Y_l\cdot Y_R$, and
$\underbrace{\cdot\;\cdot} $ denotes the propagation of
transversal U(1)-gauge field $B_{\mu}^{\perp}
=(g_{\mu\nu}-\partial_\mu\partial_\nu/\partial^2)B^{\nu}$, which
is independent of the longitudinal mode, so that
$$
\underbrace{B_\mu^{\perp}(x) B_\nu^{\perp}(0)} = -i g_{\mu\nu}\int
\frac{d^4 p}{(2\pi)^4} e^{ipx} \frac{1}{p^2}
$$
to the leading order of perturbative theory. To the bare order the
equation of motion for the bi-local field results in the
straightforward substitution of local field $\phi$, as it stands
in the above consideration for the correlators, developing the
vacuum expectation values. After the analysis of divergences in
the $J$-dependent Green functions, the corresponding contra-terms
must be added to the action. Then the $J$-source can be integrated
out or renormalized, that results in a Higgs-like action,
containing some couplings to fermions as well as a suitable
potential to develop the spontaneous breaking of electroweak
symmetry.

We stress that there are no other suitable composite operators appearing in the
second order of SM gauge symmetry with the quantum numbers relevant to the
Higgs interactions providing the generation of fermion masses through the
Yukawa-like couplings except the operators described above.

In this paper we calculate the effective potential up to the quartic term for
the sources corresponding to the bi-local composite operators of quarks and
leptons to the one-loop accuracy of renormalization in the SM. The
normalization condition of potential parameters: $\mu^2$ and $\lambda$ standing
in
$$
V(J^\dagger,J) = -\mu^2\cdot J^\dagger J+\lambda \cdot (J^\dagger J)^2,
$$
is strictly defined in the SM, since we do not involve some additional
interactions. Therefore, both $\mu^2$ and $\lambda$ for a nonfundamental source
must be equal to zero, exactly, i.e. $V=0$, which, however, can be satisfied at
a single scale $M$ because of logarithmic renormalization for couplings, so
that
\begin{equation}
\mu^2(M) =0,\;\;\; \lambda(M)=0.
\label{matchnull}
\end{equation}
It is essential that the choice of composite operators is conformed to the
effective action of SM in the second order over the gauge couplings. Otherwise,
the introduction of arbitrary composite operators with the given properties
with respect to the gauge symmetry generally does not imply the imposition of
matching condition in (\ref{matchnull}), which is extremely important, since it
removes an uncertainty of the potential due to a finite renormalization of
parameters.

Below $M$, the mass parameter $\mu^2(\Lambda)$ depending on the ``infrared
cut-off $\Lambda$, is positive, and the electroweak symmetry is broken down.
So, we suppose that the bi-local representation is valid in the range of
virtualities: $[\Lambda;M]$, and below $\Lambda$ we can explore the local
Higgs fields.

As was shown in ref. \cite{HaHa}, a variety of composite operators appropriate
for the Higgs quantum numbers can be rearranged so that practically arbitrary
values of higgs mass or $t$ quark mass could be derived. In other words, in
order to get a definite description of higgs sector, one should to suppress
contributions by a lot of composite operators except the special ones. this
dominance of several composite operators is usually motivated by an extended
dynamics beyond the SM, for instance, by the thechnicolor providing the
dominance of some bound channels.

In the present paper, the choice of dominant composite operators is dictated by
the SM gauge symmetry, since we isolate the only composite structure in the
second order over the gauge couplings, while the appearance of other operators
takes place at higher orders, and, hence, their contributions must be
suppressed\footnote{This suppression becomes even better at higher virtualities
because of the asymptotic freedom of nonabelian interactions, while the abelian
charge remains small up to the GUT scale.}. Moreover, this motivation on the
form of composite operators makes us to add the matching condition of
(\ref{matchnull}), which is a new idea for the composite models, and it is
certainly due to the electroweak nature of composite operators.

Thus, for a walker travelling from low scales to  higher ones, the
whole picture of electroweak symmetry breaking looks as the
following:
\begin{itemize}
\item[1.] The SM extension with several Higgs fields is the local theory with
the ultraviolet cut-off $\Lambda$.
\item[2.] The parameters of Higgs potential at the scale $\Lambda$ is matched
with the effective potential of bi-local source, calculated in the range
$[\Lambda; M]$, so that $M$ denotes the scale, where the potential is exactly
zero.
\end{itemize}
The value of $\Lambda$, hence, can be related with the masses of
gauge bosons, or the vacuum expectation value ({\sc vev}) $v_{\rm
SM}$ for the Higgs field in the SM. The value of $M$ with respect
to $\Lambda$ is fixed by two simple requirements: at the matching
point $\Lambda$ the Yukawa constant of $t$-quark calculated with
the composite operators is determined by the condition of infrared
fixed point in the local theory \cite{Hill}, while the Yukawa
constant is expressed in terms of abelian gauge coupling at the
scales of $\Lambda$ and $M$ due to the consistent matching
condition of potential (\ref{matchnull}). At fixed $\Lambda$ this
supposition makes $M$ to grow to the GUT scale, that implies the
solution of naturalness. So, we can read off the third point:
\begin{itemize}
\item[3.] The Yukawa constants of heaviest fermions in the local theory have
the matching conditions at $\Lambda$ to the couplings given by the bi-local
representation, so that the infrared fixed point for the $t$-quarks is exactly
reached.
\end{itemize}
The masses of $b$-quark and $\tau$-lepton can be also calculated after the use
of both the definite matching at $\Lambda$ and infrared fixed points in the RG
equations below $\Lambda$.

Finally, the potential of Higgs fields at $\Lambda$ can serve to estimate the
masses of neutral scalar particles by means of RG evolution and the infrared
fixed point for the quartic vertex $\lambda$.
The important property of fixed points under consideration is that the Yukawa
constants and quartic coupling are given by appropriate combinations of gauge
coupling constants.

Thus, the local theory with the local Higgs fields and the electroweak symmetry
breaking can be certainly matched to the effective potential of sources for the
bi-local composite operators of quarks and leptons at the scale $\Lambda$ and
to the corresponding Yukawa constants, which are calculable in the region of
virtualities $[\Lambda; M]$, so that the fixed point matching of $t$-quark
coupling and the symmetry matching-condition of null effective potential
(\ref{matchnull}) result in $M$ living in the GUT area.

Then we find the following general results:

1. Three bi-local composite operators formed by the fermions of heaviest
generation, develop the effective potential of their sources, so that nonzero
{\sc vev}'s break the electroweak symmetry. We treat these dynamics above the
scale $\Lambda$ as the strong self-interaction regime for three independent
scalar higgses as equivalent to the weak self-interaction regime for three
independent sources of composite operators.

2. The position of matching point $\Lambda=633$ GeV is fixed by the measured
masses of gauge bosons, after the higgs sector is given by three independent
scalar fields.

3. At $\Lambda$ the infrared fixed point condition is satisfied for the
Yukawa coupling of $t$ quark, only, while the couplings of $b$ quark and $\tau$
lepton evolve to the fixed points at lower scales in agreement with the current
data available. The masses of higgses evolve too.

4. Under the item 3, the position of ultraviolet cut off $M\sim
10^{12}-10^{19}$ GeV with respect to $\Lambda$ is given by the condition of
zero effective potential for the sources of composite operators, that is
governed by the renormalization group for the ${\sf U}(1)$ hypercharge, so that
we find the natural hierarchy for $M$ and $\Lambda$.

The paper is organized in the following way: Section II is devoted to the
definition of sources for the bi-local composite operators and calculation of
effective potential to the one-loop accuracy. The masses of gauge bosons and
Yukawa constants of fermions are evaluated in Section III at the scale
$\Lambda$. The exploration of infrared fixed point conditions for the Yukawa
constants and quartic Higgs coupling is considered in Section IV. Numerical
estimates of masses for the heaviest fermions as well as the Higgs fields are
given in Section V. In Section VI we shortly discuss the problem of generations
and the vacuum structure. The obtained results and the points of discussion are
summarized in Conclusion.

\section{Sources of composite operators and effective potential}

Let us define the following bare actions for the sources of bi-local operators
\begin{eqnarray}
S_\tau &=& \int dx dy\; N_\tau\cdot J_{\tau}^\dagger (x,y)\; [\bar
\tau_R(x)\; \underbrace{\slashchar{B}^{\perp}(x)
\slashchar{B}^{\perp}(y)}
\; \tau_L(y)] +{\rm h.c.},\nonumber \\
S_t &=& \int dx dy\; N_t\cdot J_t^\dagger (x,y)\; [\bar
t_R(x)\cdot n\; \underbrace{\slashchar{B}^{\perp}(x)
\slashchar{B}^{\perp}(y)}
\; \bar n\cdot t_L(y)] +{\rm h.c.}, \label{act}\\
S_b &=& \int dx dy\; N_b\cdot J_b^\dagger (x,y)\; [\bar
b_R(x)\cdot n\; \underbrace{\slashchar{B}^{\perp}(x)
\slashchar{B}^{\perp}(y)} \; \bar n\cdot b_L(y)] +{\rm
h.c.},\nonumber
\end{eqnarray}
where we have introduced the ${\sf SU(3)}$-triplet unit-vector $n_i$, so that
$\bar n\cdot n=1$, and the $n$-dependent terms in the effective action after
the account for the loop-corrections have to be averaged over $n_i$ to restore
the explicit invariance under the transformations of ${\sf SU(3)}$. For
instance, since we generally have $n_i\bar n_j= \frac{1}{3} \delta_{ij}+
\frac{1}{\sqrt{3}}\lambda^a_{ij} F_a$, we can straightforwardly check that
$$
\langle n_i \bar n_j\rangle = \frac{1}{3}\delta_{ij},\;\;\;
\langle F_a\rangle=0,\;\;\; \langle F_a F_b\rangle = \frac{1}{8}\delta_{ab},
$$
and so on.

For nonzero $Y_L$ and $Y_R$, which are under consideration, we can redefine
the factors $N_J$ to include the hypercharges into the definition of sources,
so that $\tilde N_p = \alpha_Y$ and $\tilde J_p =\pi\cdot Y_L \cdot Y_R J_p$
for $p=t,\; b,\; \tau$, which will not change the final results concerning for
the physical quantities: masses and couplings. In what follows we will omit the
tildes for the sake of briefness.

\setlength{\unitlength}{1mm}
\begin{figure}[th]
\begin{center}
\begin{picture}(70,40)
\put(7,0){\epsfxsize=5.5cm \epsfbox{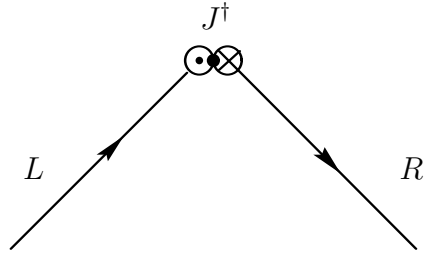}}
\put(33.5,30){$J^\dagger$}
\put(10,10){$L$}
\put(60,10){$R$}
\end{picture}
\end{center}
\caption{The vertex of global source $J^\dagger$ for the bi-local operator of
left-handed and right-handed fermions, where the huge dot denotes the
propagation of hypercharge gauge boson.}
\label{vert}
\end{figure}

In the calculations of effective potential we consider the global values of
sources independent of local coordinates: $\partial_{x,y} J(x,y) \equiv 0$. The
corresponding vertex derived from actions (\ref{act}) is shown in Fig.
\ref{vert}.
For the $t$-quark it has the form
\begin{equation}
\Gamma_t = i\alpha_Y\; J^\dagger\; \bar t_R(p)\cdot n\; \frac{-4i}{p^2}\;
\bar n\cdot t_L(p)+{\rm h.c.}
\end{equation}

\begin{figure}[th]
\begin{center}
\begin{picture}(70,40)
\put(7,0){\epsfxsize=5.5cm \epsfbox{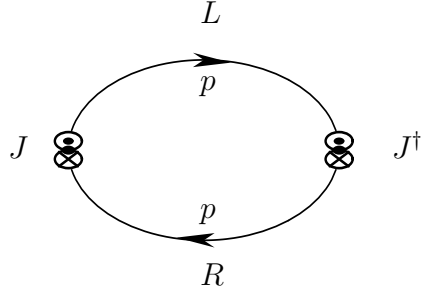}}
\put(33.5,35){$L$}
\put(33.5,0){$R$}
\put(33.5,26){$p$}
\put(33.5,9){$p$}
\put(8,17){$J$}
\put(59,17){$J^\dagger$}
\end{picture}
\end{center}
\caption{The $J^\dagger J$-term  in the effective potential.}
\label{mu}
\end{figure}

\begin{figure}[th]
\begin{center}
\begin{picture}(110,40)
\put(7,0){\epsfxsize=4cm \epsfbox{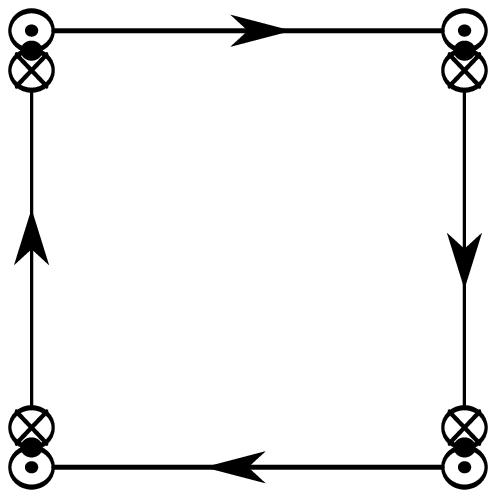}}
\put(43.5,36){$J^\dagger$}
\put(43.5,0){$J$}
\put(2,0){$J^\dagger$}
\put(2,36){$J$}
\put(2,17.5){$R$}
\put(43.5,17.5){$R$}
\put(67,0){\epsfxsize=4cm \epsfbox{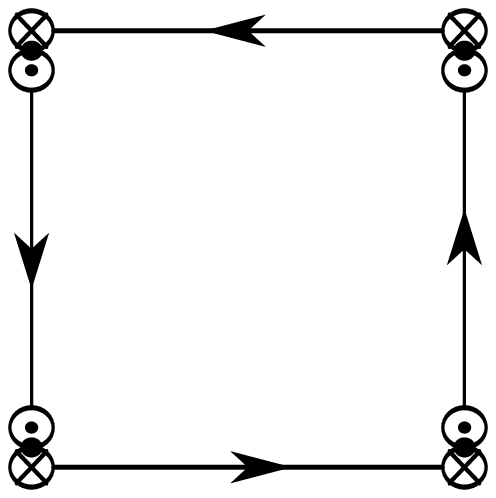}}
\put(103.5,36){$J^\dagger$}
\put(103.5,0){$J$}
\put(62,0){$J^\dagger$}
\put(62,36){$J$}
\put(62,17.5){$L$}
\put(103.5,17.5){$L$}
\end{picture}
\end{center}
\caption{The $(J^\dagger J)^2$-term in the effective potential.}
\label{lam}
\end{figure}

The diagrams for the calculation of quadratic and quartic terms of effective
potential are shown in Figs. \ref{mu} and \ref{lam}, respectively.

The parameters of potential
\begin{equation}
V(J^\dagger,J) = -\mu^2\cdot J^\dagger J+\lambda \cdot (J^\dagger J)^2,
\end{equation}
can be written down in the euclidean space as
\begin{eqnarray}
i\mu^2_B & = & -i N_J^2 \int_{\Lambda^2}^{M^2} \frac{d^4 p}{(2\pi)^4}\;
\frac{4^2{\rm tr}[P_L \slashchar{p} \slashchar{p}]}{(p^2)^4},\\
-i4\lambda_B & = & -2 i N_J^4 \int_{\Lambda^2}^{M^2} \frac{d^4 p}{(2\pi)^4}\;
\frac{4^4{\rm tr}[P_L \slashchar{p} \slashchar{p} \slashchar{p}
\slashchar{p}]}{(p^2)^8},
\end{eqnarray}
\noindent
which are independent of the fermion flavor. Here $P_L=
\frac{1}{2}(1-\gamma_5)$ is the projector on the left-handed fermions.

Supposing $M^2\gg \Lambda^2$, we find
\begin{eqnarray}
-\mu^2_B & = &  N_J^2\; \frac{2}{\pi^2}\; \frac{1}{\Lambda^2},\\
\lambda_B & = & N_J^4\; \frac{4}{\pi^2}\; \frac{1}{\Lambda^8}.
\end{eqnarray}

As we have already mentioned in the Introduction, the effective potential has
to be subtracted, so that at the scale $M$ it equals zero, exactly, since we
deal with the source of composite operators not involving some interactions
beyond the gauge interactions of SM. Then, we get
\begin{eqnarray}
\mu^2_R(\Lambda) & = &  \frac{2}{\pi^2}\; \frac{1}{\Lambda^2}\; \alpha_Y^2(M)
(1-\varkappa^2(\Lambda)),\\
\lambda_R(\Lambda) & = & \frac{4}{\pi^2}\; \frac{1}{\Lambda^8}\; \alpha_Y^4(M)
(1-\varkappa^2(\Lambda))^2,
\end{eqnarray}
where we have introduced the notation for
$$
\varkappa(\Lambda) = \frac{\alpha_Y(\Lambda)}{\alpha_Y(M)},
$$
with the normalization $\varkappa(M)=1$. The scale-independent factors
$\alpha_Y^{2,4}(M)$ can be removed by the redefinition of sources: $J^\prime =
\alpha_Y J$, which we imply below. In addition we introduce $J(\Lambda) =
\frac{1}{\Lambda^2} J^\prime$ to obtain more usual notations. Then,
\begin{eqnarray}
\mu^2(\Lambda) & = &  \frac{2}{\pi^2}\;
(1-\varkappa^2(\Lambda))\;{\Lambda^2},\\
\lambda(\Lambda) & = & \frac{4}{\pi^2}\; (1-\varkappa^2(\Lambda))^2.
\end{eqnarray}

The vacuum expectation value, {\sc vev}, is given by $\langle J^\dagger
J\rangle = \frac{\mu^2}{2\lambda}$, so that
\begin{equation}
\langle J^\dagger(\Lambda) J(\Lambda)\rangle  = \frac{1}{4}\;
\frac{1}{1-\varkappa^2(\Lambda)}\; \Lambda^2.
\end{equation}
Remember, that the potential parameters are the same for all charged heavy
fermions: $t$-quark, $b$-quark and $\tau$-lepton. The density of vacuum energy
is independent of flavor, too,
$$
V({\rm vac}) = -\frac{\mu^4}{4\lambda} = -\frac{1}{4 \pi^2}\;\Lambda^4.
$$
Then the action represented as the sum of terms over the space-time intervals
with $d^4 x \sim 1/\Lambda^4$, has the form
$$
S({\rm vac}) = -\int d^4 x\; V({\rm vac})\sim \sum \frac{1}{4\pi^2},
$$
and it is independent of $\Lambda$.

\section{Masses of gauge bosons and Yukawa constants}

\begin{figure}[th]
\begin{center}
\begin{picture}(140,60)
\put(7,0){\epsfxsize=5.5cm \epsfbox{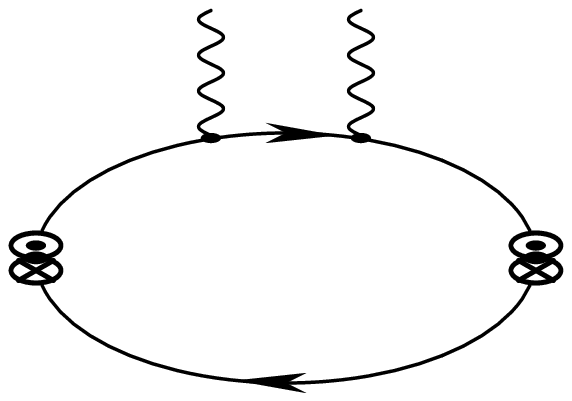}}
\put(38,39){$A_j$}
\put(28,39){$A_i$}
\put(9,17){$J$}
\put(56,17){$J^\dagger$}
\put(67,0){\epsfxsize=5.5cm \epsfbox{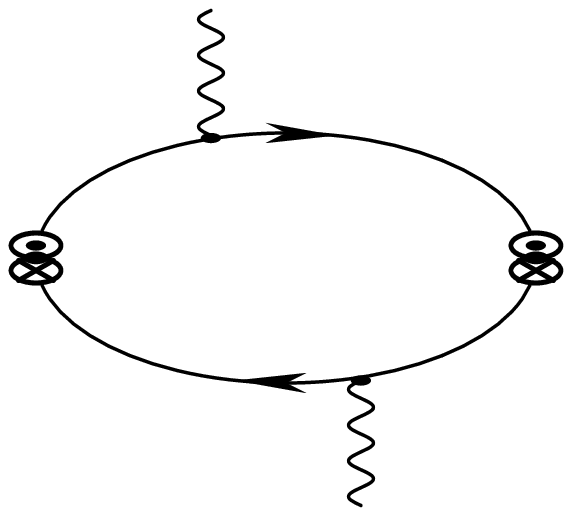}}
\put(98,-5){$A_j$}
\put(88,39){$A_i$}
\put(69,17){$J$}
\put(116,17){$J^\dagger$}
\end{picture}
\end{center}
\caption{The $(A_i A_j)$-terms in the effective potential of gauge bosons.}
\label{mw}
\end{figure}

The diagrams, which result in the masses of gauge bosons, are shown in Fig.
\ref{mw}, where the permutations over the gauge bosons are implied.

We straightforwardly find that the couplings of gauge bosons are proportional
to the differences of their charges, so that
$$
m_{12}^2 A_1^\mu A_2^\nu g_{\mu\nu} \sim (Q_1^L-Q_1^R) (Q_2^L-Q_2^R) A_1^\mu
A_2^\nu g_{\mu\nu},
$$
where $Q^{L,R}$ denote the charges of left-handed and right-handed fermions.
This implies that the vector-like gauge bosons, i.e. when $Q^L=Q^R$, remain
massless.

For the $W$- and $Z$-bosons after the subtraction procedure of $\varkappa^2 \to
(1-\varkappa^2)$, we find
\begin{eqnarray}
m_W^2 & = & \frac{4\pi \alpha_2}{2} \sum_p \frac{\Lambda_p^2}{4\pi^2},\\
m_Z^2 & = & m_W^2 \; \frac{1}{\cos^2 \theta_W},
\end{eqnarray}
where $\theta_W$ is the Weinberg angle \cite{Wein}, as usual, and the sum is
taken over the heavy flavors $p=t,\; b,\; \tau$. As we have seen in the
previous section $\Lambda_p=\Lambda$ is independent of flavor, and, hence, we
can introduce the Higgs field $h_p$ with the {\sc vev}, $\langle h_p\rangle=v$,
so that
$$
v=\frac{\Lambda}{2\pi},\;\;\;\; h_p(v) = \frac{1}{\pi}
\sqrt{1-\varkappa^2(v)}\;
J_p(v).
$$
Thus, we get
\begin{equation}
m_W^2 = \frac{4\pi \alpha_2}{2}\; 3 v^2,
\end{equation}
so that $v_{\rm SM}^2 = 3 v^2 \approx (174\; {\rm GeV})^2$, when the potential
at the scale $\Lambda$ has the form\footnote{There is a possibility to change
the convention on the prescription of scale by replacing $\Lambda\to v$.}
\begin{equation}
V(h_p,h_p^\dagger) = -2 \Lambda^2\; h_p^\dagger\cdot h_p + (2\pi)^2\;
(h_p^\dagger\cdot h_p)^2.
\end{equation}
We check that the quadratic term $-2\Lambda^2$ is exactly given by the one-loop
calculation in the local $\phi^4$-theory with $\lambda=(2\pi)^2$ and cut-off
$\Lambda$.

The masses of fermions at the same scale can be derived from the diagram shown
in Fig. \ref{vert} by putting the fermion momenta to the given virtuality,
$p^2=\Lambda^2$. Then, after the appropriate subtraction [$\varkappa\to
(1-\varkappa)$] we
get
$$
m_p = \lambda_p \cdot v,
$$
with
\begin{eqnarray}
\lambda_t(v) = \lambda_b(v) &=& \frac{4\pi}{3\sqrt{2}}\;
\sqrt{\frac{1-\varkappa(v)}{1+\varkappa(v)}},\\
\lambda_\tau(v) &=& 3 \lambda_t(v).
\end{eqnarray}
Replacing $v$ by $v_{\rm SM}$, we find $\lambda_p^{\rm SM} =
\lambda_p/\sqrt{3}$.

Thus, we have calculated the masses of gauge bosons, Yukawa constants of
heaviest fermions and the parameters of Higgs potential at the scale $\Lambda$,
which have to be matched with the quantities of local theory valid below
$\Lambda$.

\section{Infrared fixed points}
To the moment we have the local theory with three neutral Higgs fields, which
are coupled with the appropriate heavy fermions in each sector, with the
cut-off $\Lambda$, where the Yukawa couplings have to be matched with the
values calculated in the effective potential of sources for the composite
operators.

The one-loop RG equations for the couplings\footnote{The corresponding two-loop
RG equations are given in ref. \cite{Machacek:1983fi}. We shortly comment the
influence of two-loop corrections below.} have the form \cite{Pok}
\begin{eqnarray}
\frac{d\ln \lambda_t}{d\ln \mu} & = & \frac{1}{(4\pi)^2}\; \left[ \frac{9}{2}
\lambda_t^2 -8 g_3^2 -\frac{17}{12} g_Y^2 -\frac{9}{4} g_2^2\right],\nonumber\\
\frac{d\ln \lambda_b}{d\ln \mu} & = & \frac{1}{(4\pi)^2}\; \left[ \frac{9}{2}
\lambda_b^2 -8 g_3^2 -\frac{5}{12} g_Y^2 -\frac{9}{4} g_2^2\right],
\label{RG1}\\
\frac{d\ln \lambda_\tau}{d\ln \mu} & = & \frac{1}{(4\pi)^2}\; \left[
\frac{5}{2} \lambda_\tau^2 -\frac{15}{4} g_Y^2 -\frac{9}{4}
g_2^2\right],\nonumber
\end{eqnarray}
where $g_3^2 = 4\pi \alpha_s$ is the QCD coupling, $g_Y^2 = 4\pi \alpha_Y$ is
the hypercharge coupling, and $g_2^2 = 4\pi \alpha_2$ is the $\sf SU(2)$-group
coupling. At ``low'' virtualities about $v\sim 100$ GeV, the dominant
contribution to the $\beta$-functions of quark couplings is given by QCD. We
suppose that the value of matching point $\Lambda$ is dictated by the fixed
point condition for the $t$-quark: $\frac{d\ln \lambda_t}{d\ln \mu}=0$
\cite{Hill}, i.e.
\begin{equation}
\lambda_t^2(v) = \frac{64\pi}{9}\; \alpha_s(v)+\frac{34\pi}{27}\;
\alpha_Y(v)+2\pi \alpha_2(v)\approx \frac{64\pi}{9}\; \alpha_s(v),
\label{app1}
\end{equation}
when the matching gives
\begin{equation}
\lambda_t^2(v) = \frac{8\pi^2}{9}\; \frac{1-\varkappa(v)}{1+\varkappa(v)}.
\end{equation}
Therefore, we find
\begin{equation}
\varkappa(v) = \frac{1-\frac{8\alpha_s(v)}{\pi}}{1+\frac{8\alpha_s(v)}{\pi}}.
\end{equation}
Due to the contribution by the hypercharge the difference between the RG
equations for $\lambda_b$ and $\lambda_t$ causes the reach of infrared fixed
point for the $b$-quark at a lower scale than for the $t$-quark. Indeed, the
fixed point condition for the $b$-quark reads off
\begin{equation}
\frac{9}{2}(\lambda_b^2(\mu)-\lambda_t^2(\mu)) = -g_Y^2(\mu).
\label{fb}
\end{equation}
Making use of matching condition $\lambda_b(v)=\lambda_t(v)$, we can write down
$$
\frac{d\ln \lambda_t/\lambda_b}{d\ln \mu}= - \frac{1}{(4\pi)^2}\; g_Y^2,
$$
for small changes, so that
\begin{equation}
\lambda_b(\mu)-\lambda_t(\mu)\approx -\frac{\lambda_t(\mu)}{(4\pi)^2}\; g_Y^2\;
\ln\frac{\Lambda}{\mu}.
\label{rb}
\end{equation}
Then, we can derive from (\ref{fb}) and (\ref{rb}) the following estimate of
current mass for the $b$-quark
\begin{equation}
\ln\frac{m_t}{m_b(\hat v_b)} = \frac{\pi}{4\alpha_s(m_b(\hat v_b))},
\end{equation}
where the current mass of $t$-quark is given by
$$
m_t(m_t) = \frac{8}{3} \sqrt{\pi \alpha_s(v)}\cdot v,
$$
since the evolution of $t$-quark mass above the scale $v$ is determined by the
running of effective constant, which is negligibly small in the interval
$[v,m_t]$, and, hence, $m_t(m_t)\approx m_t(v)$ with quite a high accuracy.
The scale of $b$-quark normalization is given by the following
$$
m_b(\hat v_b) = \frac{8}{3}\sqrt{\pi\alpha_s(\hat v_b)}\cdot \hat v_b,
$$
and we use the QCD evolution to extract the current mass of $b$-quark at the
scale of its value
$$
m_b(m_b) = m_b(\hat v_b)\left(\frac{\alpha_s(m_b(m_b))}{\alpha_s(\hat
v_b)}\right)^{12/25}.
$$

Next, we can evaluate the mass of $\tau$-lepton in the same manner. At low
energies we modify the RG equation for the $\tau$-coupling, neglecting the
four-fermion weak interactions and taking into account the photon contribution.
So, we have
\begin{equation}
\frac{d\ln \lambda_\tau}{d\ln \mu}  =  \frac{1}{(4\pi)^2}\; \left[
\frac{5}{2} \lambda_\tau^2 -24\pi \alpha_{em}\right],
\end{equation}
and the infrared fixed point condition reads off
\begin{equation}
\lambda_\tau^2 = \frac{48\pi}{5}\; \alpha_{em}.
\label{ftau}
\end{equation}
The change of $\lambda_\tau$ from the matching value
$\lambda_\tau^2=9\lambda_t^2 = 64\pi\alpha_s$ can be found in the solution of
\begin{equation}
\frac{d\ln \lambda_\tau}{d\ln \mu}  \approx  \frac{1}{(4\pi)^2}\;
\frac{5}{2} \cdot 9 \lambda_t^2 \approx 40\; \frac{\alpha_s}{4\pi},
\label{tauev}
\end{equation}
so that
\begin{equation}
\lambda_\tau(\mu) =
\lambda_\tau(v)\cdot\left(\frac{\alpha_s(\mu)}
{\alpha_s(v)}\right)^{-\frac{40}{2 b_3}},
\label{rtau}
\end{equation}
where $b_3 = 11-\frac{2}{3}n_f =9$ at $n_f=3$. From (\ref{ftau}), (\ref{rtau})
and $\lambda_t^2= 64\pi \alpha_s/9$ we deduce the relation
\begin{equation}
\alpha_s(m_\tau) =
\alpha_s(v)\cdot\left(\frac{3}{20}\;\frac{\alpha_{em}(m_\tau)}{\alpha_s(v)}
\right)^{-\frac{9}{40}}.
\end{equation}
Note, that the one-loop evolution to such the large change of scales is quite a
rough approximation. To improve the estimate of $\tau$-lepton mass we integrate
(\ref{tauev}) numerically with the same boundary conditions and extract the
value under consideration.

Let us consider the way to estimate the masses of neutral Higgs bosons. The RG
equations for the quartic couplings of scalar particles with the heaviest
fermions are represented by the following:
\begin{equation}
\frac{d \lambda}{d\ln\mu} = \frac{3}{2\pi^2}\left[\lambda^2-\frac{a_p}{4}\;
\lambda_p^4\right],
\end{equation}
where $a_t=a_b=1$, $a_\tau=\frac{1}{3}$, and we neglect the contribution given
by the electroweak gauge couplings. This approximation is quite reasonable,
since at $\Lambda(v)$ the quartic couplings $\lambda(v)=(2\pi)^2$ dominate. For
the Higgs fields coupled to the $t$- and $b$-quarks, the infrared fixed points
coincide with each other to the order under consideration, so that
$$
\lambda(\mu_H) \approx \frac{1}{2}\; \lambda^2_{t,b}(\mu_H) = \frac{32\pi}{9}\;
\alpha_s(\mu_H),
$$
which implies that the corresponding masses of scalars are degenerated with a
high accuracy. Let us evaluate the scale of reaching the infrared fixed point.
The evolution can be approximated at large $\lambda$ by the equation
$$
\frac{1}{\lambda(\mu_H)} =
\frac{1}{\lambda(v)}+\frac{3}{2\pi^2}\;\ln\frac{v}{\mu_H},
$$
so that we derive
\begin{equation}
\ln \frac{v}{\mu_H} \approx
\frac{3\pi}{16\alpha_s(\mu_H)}-\frac{1}{6},
\label{str-h}
\end{equation}
If we use the RG evolution for the QCD
coupling \cite{PEP}
$$
\frac{1}{\alpha_s(\mu_H)} = \frac{1}{\alpha_s(v)}-
\frac{b_3}{2\pi}\; \ln\frac{v}{\mu_H},
$$
at $n_f=5$, we arrive to
\begin{equation}
\ln \frac{v}{\mu_H} \approx \frac{6}{55}\;\frac{\pi}{\alpha_s(v)},
\end{equation}
although the straightforward equation for the scale in (\ref{str-h}) can be
more accurate numerically.

Thus, following the general relation for the mass of Higgs field,
$$
m_H(\mu) = 2 \sqrt{\lambda(\mu)}\cdot v,
$$
we have the estimates
\begin{eqnarray}
m_H(v)     &=& 4\pi\cdot v,\\
m_H(\mu_H) &=& \frac{8}{3}\sqrt{2\pi\alpha_s(\mu_H)} \cdot v.
\end{eqnarray}
As for the Higgs field coupled to the $\tau$-lepton, it is quite easily
recognize that the corresponding scale $\mu$ is much greater than for the
scalars coupled with the heaviest quarks, and, hence, its mass is greater than
we have considered above. Indeed, we can use the evolution of $\lambda_\tau$ at
large scales, where it is driven as $\lambda_\tau = 3 \lambda_t$, so that we
derive the relation analogous to (\ref{str-h})
\begin{equation}
\ln \frac{v}{\mu_{H_\tau}} \approx
\frac{\pi}{16\sqrt{3}\alpha_s(\mu_{H_\tau})}-\frac{1}{6},
\label{str-htau}
\end{equation}
and
$$
m_{H_\tau}(\mu_{H_\tau}) =
{8}\sqrt{\frac{2}{\sqrt{3}}\pi\alpha_s(\mu_{H_\tau})} \cdot v.
$$
To the moment we are ready to get numerical estimates.

\section{Numerical evaluation and the naturalness}

First of all, the {\sc vev}'s of Higgs fields are directly given by the masses
of gauge bosons, so that
$$
v = 100.8\pm 0.1\; {\rm GeV},
$$
and the cut-off
$$
\Lambda = 2\pi v= 633.0\pm 0.6\; {\rm GeV,}
$$
where we use the experimental data shown in Table \ref{tab}.

\begin{table}[th]
\begin{center}
\begin{tabular}{||p{2cm}|p{2.5cm}||}
\hline
$m_W$, GeV      & $80.41\pm 0.09 $ \\
\hline
$\alpha_2^{-1}$ & $29.60\pm 0.04 $ \\
\hline
$m_t$, GeV      & $174\pm 5      $ \\
%$\alpha_s(m_Z)$ & $0.119\pm 0.002$ \\
\hline
\end{tabular}
\end{center}
\caption{The experimental data on the electroweak parameters
[5,6].} % \cite{SM,PEP}.}
\label{tab}
\end{table}

The estimates for the masses of fermions depend on the values of QCD coupling
constant. We put the value\footnote{The central value is slightly displaced
from the ``world average'' $\alpha_s(m_Z)=0.119\pm 0.002$ \cite{PDG}, though it
is within the current uncertainty. However, this parameter corresponds to the
LEP fit \cite{SM} as well as to the recent global fit of structure functions
\cite{AK}.}
$$
\alpha_s(m_Z)=0.122\pm 0.003,
$$
which corresponds to the $\Lambda^{(5)}_{\overline{\rm MS}}=255\pm 45$ MeV in
the three-loop approximation for the $\beta$-function. We suppose that the
threshold values for the changing the number of active quark flavors are equal
to $\hat m_b=4.3$ GeV and $\hat m_c=1.3$ GeV. The variation of threshold values
is not so important in the estimates in contrast to the uncertainty in
$\alpha_s$, which dominates in the error-bars.

Then we can numerically solve the equations in the previous section to find the
current masses
\begin{eqnarray}
m_t(m_t) & = & 165\pm 1\; {\rm GeV,} \nonumber\\
m_b(m_b) & = & 4.18\pm 0.38\; {\rm GeV,} \\
m_\tau(m_\tau) & = & 1.78\pm 0.27\; {\rm GeV.}\nonumber
\end{eqnarray}
The one-loop relation of perturbative QCD for the pole mass of quark \cite{PEP}
is given by
$$
m^{(p)} = m(m) \left(1+\frac{4}{3\pi}\; \alpha_s(m)\right).
$$
Then we estimate
\begin{eqnarray}
m_t^{(p)} &=& 173\pm 2\; {\rm GeV},\nonumber\\
m_b^{(p)} &=& 4.62\pm 0.40\; {\rm GeV}.\nonumber
\end{eqnarray}
The QED correction to the $\tau$-lepton mass is negligibly small.

We see that the $t$-quark mass is in a good agreement with the direct
measurements. The $b$-quark mass is in the desirable region. It is close to
that of estimated in the QCD sum rules \cite{SVZ}, where $m_b(m_b)=4.25\pm
0.15$ GeV \cite{BM}, and in the potential approach \cite{KKO}, where
$m_b(m_b)=4.20\pm 0.06$ GeV. It is worth to note that the pole mass is not the
value, which has a good convergency in the OPE approach (see references in
\cite{BM,KKO}), so we present it to the first order for the sake of reference.
However, we stress also that the deviations from the central values are caused
by the uncertainties in the $\alpha_s$ running.

The infrared fixed masses of neutral scalars, coupled with the $t$- and
$b$-quarks and the $\tau$-lepton, equal
\begin{equation}
m_H = 306\pm 5\; {\rm GeV,}\;\;\;\; m_{H_\tau} = 552\pm 9\; {\rm GeV,}
\label{hg}
\end{equation}
which can be compared with the global fit of SM at LEP yielding $m_H=
76^{+85}_{-47}$ GeV \cite{SM}. The central value of this fit was recently
excluded by the direct searches at modern LEP energies, where the constraint
was obtained $m_H > 95$ GeV \cite{SM,PEP}. We expect, however, that
many-doublet models of Higgs sector have a different connection to the LEP
data. Indeed, the fit of SM with the single Higgs particle yields the value for
the logarithm $l_H=\log_{10} m_H^{\rm SM}[{\rm GeV}] = 1.88^{+0.33}_{-0.41}$,
whereas this correction basically contributes into the observed quantities due
to the coupling to the massive gauge bosons. Then, we can write down the
following approximation for this value in the model under consideration:
$$
l_H = \frac{1}{3}\sum_p \kappa_p \log_{10}
m_{H_p}[{\rm GeV}],
$$
where the factor $\frac{1}{3}$ represents the fraction of scalar coupling in
the squares of gauge boson masses, respectively for $p=t,\; b,\; \tau$, and
$\kappa_p$ stands for the possible formfactors at high virtualities of the
order of masses of Higgs fields. To test, we put the simple approximation
$$
\kappa_p \approx \frac{1}{1+\frac{m^2_{H_p}}{\Lambda^2}},
$$
which results in $\kappa_t=\kappa_b$ close to unit, and $\kappa_\tau\sim
\frac{1}{4}$, so that the value under consideration is equal to
$$
l_H \approx 1.86,
$$
that is optimistically close to what was observed at LEP. So, the values in
(\ref{hg}) are not in contradiction with the current data.

Next, since we deal with the strongly coupled version of Higgs sector
(remember, that $m_H(v)\approx 1267$ GeV), we need more careful consideration
of effective potential to take into account the higher dimensional operators,
representing the multi-higgs couplings. So, we keep (\ref{hg}) as soft
estimates of masses for the Higgs fields, which implies that the decays into
the massive gauge bosons are the dominant modes for these scalar particles .

Finally, we evaluate the scale $M$, where the electroweak symmetry has to be
exactly restored. The value of $\varkappa(v)$ is equal to
$$
\varkappa(v) = \frac{\alpha_Y(v)}{\alpha_Y(M)} = 0.532\pm 0.005,
$$
which implies $\alpha_1^{-1}(M) \approx 32$. The implication of $\varkappa$ for
$M$ depends on the running of $\alpha_Y=\frac{3}{5}\alpha_1$ \cite{PEP}:
$$
\frac{1}{\alpha_1(M)} = \frac{1}{\alpha_1(v)}+
\frac{b_1}{2\pi}\; \ln\frac{M}{v},
$$
where $b_1$ is model-dependent. So, in the SM $b_1=-\frac{4}{3}\;
n_g-\frac{1}{10}\; n_h$ with $n_g=3$ being the number of fermion generations,
$n_h$ is the number of Higgs doublets, we obtain\footnote{Numerically, we put
$\alpha_1^{-1} = 58.6$ for the order-of-magnitude estimate.}
$$
M_{\rm SM} \approx 2.5\cdot 10^{19}\; {\rm GeV,}
$$
when in the SUSY extension $b_1=-2\; n_g-\frac{3}{10}\; n_h$, so that
$$
M_{\rm SUSY} \approx 7\cdot 10^{12}\; {\rm GeV.}
$$
Hence, we obtain the broad constraints
$$
M = 7\cdot 10^{12}-2.5\cdot 10^{19}\; {\rm GeV,}
$$
and the value strongly depends on the set of fields in the region above the
cut-off $\Lambda$. At present, we cannot strictly draw a conclusion on a
preferable point. However, we can state that the offered mechanism for the
breakdown of the electroweak symmetry solves the problem of naturalness, since
the observed ``low'' scale of gauge boson masses is reasonably related to the
``high'' scale of GUT or even Planck mass.

Finally, we comment on possible uncertainties of numerical
estimates and a role of two-loop corrections. First, we analyze
the subleading terms in eq.(\ref{app1}). The gauge charges
neglected in the fixed point condition of (\ref{app1}) result in
the displacement of $t$ quark mass by a value about 4 GeV, if we
do not change the normalization of QCD coupling constant. In this
way, we note that under the account of gauge charge corrections in
(\ref{app1}) the same central value of $t$ quark mass, i.e. 165
GeV, is reproduced at $\alpha_s(m_Z^2) =0.118$, which coincides
with the Particle Data Group ``world-average''. Next, the two-loop
corrections in the RG equations for the Yukawa couplings
\cite{Machacek:1983fi} as applied to the $t$ quark lead to an
additional displacement of fixed point value. However, in this
case we have to take into account the one-loop correction to the
relation between the current mass and the pole mass of $t$ quark
due to the Higgs sector, that results in the following additive
renormalization of $m_t(m_t)$ \cite{HK}
$$
\frac{\delta m_t(m_t)}{m_t(m_t)} = - \frac{1}{16\pi^2}\,\frac{9}{2}\,
\frac{m_t^2}{v_{\rm SM}^2},
$$
at the higgs mass $m_H\approx 2 m_t$. The above correction compensates the
displacement due to the two-loop modification of infrared fixed-point condition
for the $t$ quark.

\begin{figure}[th]
\begin{center}
\begin{picture}(150,50)
\put(0,0){\epsfxsize=7cm \epsfbox{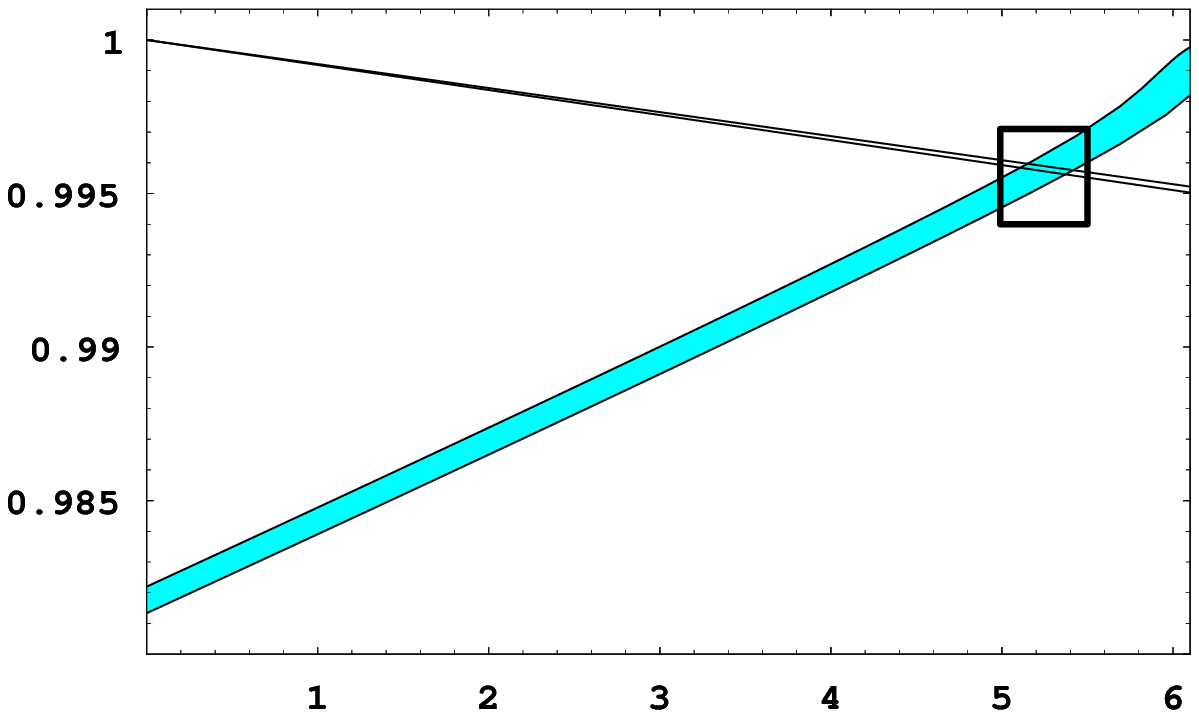}}
\put(80,0){\epsfxsize=7cm \epsfbox{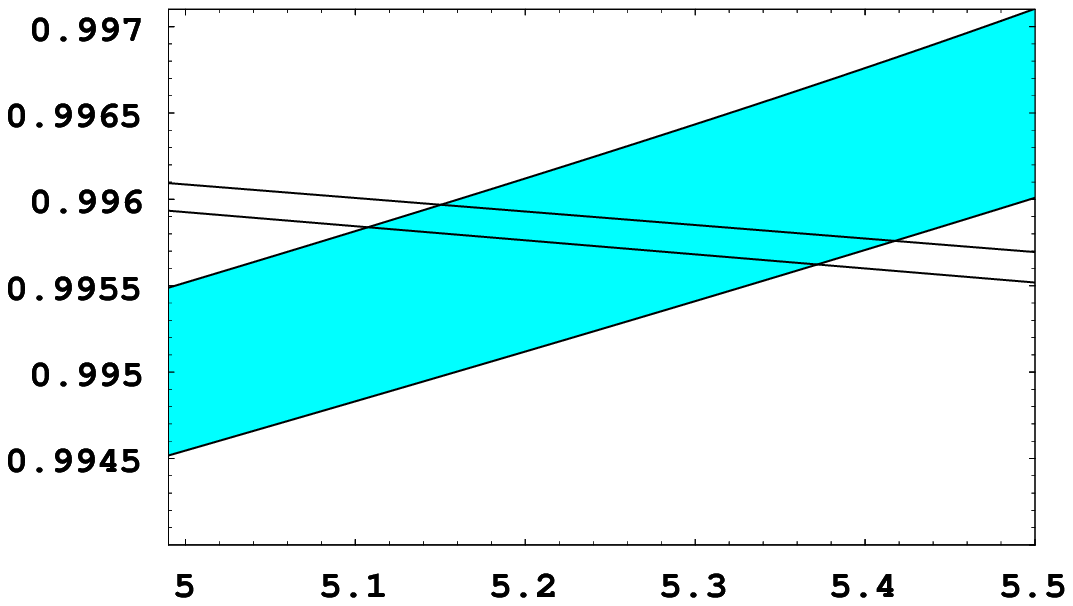}}
\put(0,45){$\lambda_b/\lambda_t$}
\put(80,45){$\lambda_b/\lambda_t$}
\put(63,-2){$\ln m_t/\mu$}
\put(140,-2){$\ln m_t/\mu$}
\end{picture}
\end{center}
\caption{The variation of RG solution for the $b$ quark Yukawa constant under
the introduction of two-loop corrections with respect to the one-loop result
(solid lines) and the uncertainty caused by the change in the normalization of
$\alpha_s$ with $\Lambda_{\overline{\rm MS}}^{(5)} = 200$ MeV and
$\Lambda_{\overline{\rm MS}}^{(5)} = 280$ MeV (the band). At the right figure
we scale the square region marked in the left picture.}
\label{uncert}
\end{figure}

Second, we study the two-loop corrections to the fixed point condition for the
$b$ quark. The corresponding modification of (\ref{fb}) reads off
\begin{equation}
\frac{9}{2}(\lambda_b^2(\mu)-\lambda_t^2(\mu)) \approx -g_Y^2(\mu)\left(1 -
\frac{1}{16\pi^2}\,\frac{4}{3}\,\lambda_t^2(\mu)\right)+O(g^4),
\label{fb2}
\end{equation}
that results in the appropriate correction in (\ref{rb}), {\it
viz.}, in the small change of slope in front of log about 2\%. The
solution of equations for the running $b$ quark mass under the
variations caused by the introduction of two-loop corrections and
the uncertainty in the coupling constant of QCD is shown in Fig.
\ref{uncert}. We can straightforwardly see that the variation of
slope in the RG equation for the Yukawa constant of $b$ quark due
to the two-loop corrections results in uncertainties, which are
much less than the variation of $b$ quark mass caused by the
uncertainties in the running coupling constant of QCD at moderate
virtualities about the $b$ quark mass. Therefore, the dominant
origin of uncertainty for the $b$ quark mass is the normalization
of $\alpha_s$. The same conclusion can be drawn for the mass of
$\tau$ lepton. Therefore, the uncertainty in the estimates of
masses for the $b$ quark and the $\tau$ lepton is not essentially
changed by the introduction of two-loop corrections, while the
value of $t$ quark mass depends on the normalization of $\alpha_s$
as well as the two-loop corrections combined, so that the
uncertainty in the current mass can reach 4 GeV in $m_t$.

As for the estimates of masses for the scalar fields, we emphasize that they
give preliminary results, and further investigations are in progress, since we
should, first, sum up subleading terms with higher powers of the higgs field
squared in the effective potential and, second, consider complete RG equations
for the quartic self-coupling, including suppressed terms. Nevertheless, our
preliminary estimates show that the scalar fields should be significantly
heavy.

\section{Generations, the number of Higgs fields and vacuum}

In the previous sections we have introduced three independent global sources
for the bi-local operators composed by the fermions of the heaviest generation,
i.e., $t$ quark, $b$ quark and $\tau$ lepton. In the SSIR, these sources
acquire the effective potentials providing the spontaneous breaking of
electroweak symmetry. Below the scale $\Lambda$ we assume the connection of
such the potentials with the potentials of local Higgs fields. Thus, we suppose
the introduction of three independent local Higgs doublets at the low energies.
Therefore, we suggest the condensation of sources related with the heaviest
generation only.

We have found the `democratic' form for the potentials of independent sources
in the SSIR. All three potentials have the same values of quadratic and quartic
couplings, while we suggest evidently broken `democracy' for the fermion
generations, since we do not introduce the condensation of sources for the
composite operators built of junior fermions.

In this section we describe a possible development on the problem of fermion
generations and the structure of vacuum in the Higgs sector.

So, let us introduce the notation of normalized {\sc vev}'s for the global
sources connected with the Higgs fields as follows:
$$
\chi_p = h_p/v, \;\;\; p=\tau,\, t,\, b,
$$
and the corresponding vacua $| 0_p\rangle$, so that
\begin{equation}
\langle 0_p| \chi_{p'} |0_{p''}\rangle = \delta_{pp'}\delta_{p'p''}.
\label{norm}
\end{equation}
Then we easily find that the mass terms
\begin{eqnarray}
{\cal L}^\tau_{Y} & \sim & \bar \tau_R \tau_L \cdot \chi_\tau + {\rm h.c.},
\nonumber \\
{\cal L}^t_{Y} & \sim & \bar t_R t_L \cdot \chi_t
+ {\rm h.c.}, \\
{\cal L}^b_{Y} & \sim & \bar b_R b_L \cdot \chi_b
+ {\rm h.c.}, \nonumber
\end{eqnarray}
could be represented by means of fields
\begin{equation}
\left(\begin{array}{c} \varphi_1 \\ \varphi_2\\ \varphi_3\end{array}\right) =
\frac{1}{\sqrt{3}}\;
\left(\begin{array}{ccc} 1 & 1 & 1 \\ 1 & \omega & \omega^2 \\ 1 & \omega^2 &
\omega \end{array}\right)
\cdot
\left(\begin{array}{c} \chi_\tau \\ \chi_t\\ \chi_b\end{array}\right)
 = U \cdot
\left(\begin{array}{c} \chi_\tau \\ \chi_t\\ \chi_b\end{array}\right) ,
\label{trans}
\end{equation}
as follows
\begin{eqnarray}
{\cal L}^\tau_{Y} & \sim & \bar \tau_R \tau_L \cdot
(\varphi_1+\varphi_2+\varphi_3) +
{\rm h.c.},
\nonumber \\
{\cal L}^t_{Y} & \sim & \bar t_R t_L \cdot (\varphi_1+\omega^2 \varphi_2+\omega
\varphi_3)
+ {\rm h.c.}, \label{generation}\\
{\cal L}^b_{Y} & \sim & \bar b_R b_L \cdot (\varphi_1+\omega \varphi_2+\omega^2
\varphi_3)
+ {\rm h.c.}, \nonumber
\end{eqnarray}
where we omit the Yukawa couplings and use the matrix $U$ defined in terms of
$\omega = \exp(i\frac{2\pi}{3})$.

Such the transformation in (\ref{trans}) relates the `heavy' basis of $\chi_p$
with the `democratic' basis of $\varphi_i$. So, the definition (\ref{trans})
can be equivalently changed by permutations of $\chi_\tau \leftrightarrow
\chi_t \leftrightarrow \chi_b$ or permutations of columns in the matrix $U$.
Such the permutations correspond to the finite cyclic group ${\mathbb Z}_3$
with the
basis $\omega$, so that complex phases of $\varphi_i$ are given by
$e^{iq\frac{2\pi}{3}}$ with the charges $q=(0,-1,1)$ of
$(\varphi_1,\varphi_2,\varphi_3)$.

Further, we can note that the vacuum fields have simple connections as follows:
\begin{eqnarray}
\left(\begin{array}{c} \varphi_1 \\ \varphi_2\\ \varphi_3\end{array}\right) =
\frac{1}{\sqrt{3}}
\left(\begin{array}{l} 1^{~} \\ 1\\ 1\end{array}\right) & \Rightarrow &
\langle 0_\tau| \boldsymbol \chi_1 |0_\tau \rangle =
\langle 0_\tau|
\left(\begin{array}{c} \chi_\tau \\ \chi_t\\ \chi_b\end{array}\right)
|0_\tau \rangle =
\left(\begin{array}{c} 1 \\ 0\\ 0\end{array}\right), \nonumber\\[4mm]
\left(\begin{array}{c} \varphi_1 \\ \varphi_2\\ \varphi_3\end{array}\right) =
\frac{1}{\sqrt{3}}
\left(\begin{array}{l} 1 \\ \omega\\ \omega^2\end{array}\right) & \Rightarrow &
\langle 0_t| \boldsymbol \chi_2 |0_t \rangle =
\langle 0_t|
\left(\begin{array}{c} \chi_\tau \\ \chi_t\\ \chi_b\end{array}\right)
|0_t \rangle=
\left(\begin{array}{c} 0 \\ 1\\ 0\end{array}\right), \\[4mm]
\left(\begin{array}{c} \varphi_1 \\ \varphi_2\\ \varphi_3\end{array}\right) =
\frac{1}{\sqrt{3}}
\left(\begin{array}{l} 1 \\ \omega^2\\ \omega\end{array}\right) & \Rightarrow &
\langle 0_b| \boldsymbol \chi_3 |0_b \rangle =
\langle 0_b|
\left(\begin{array}{c} \chi_\tau \\ \chi_t\\ \chi_b\end{array}\right)
|0_b \rangle =
\left(\begin{array}{c} 0 \\ 0\\ 1\end{array}\right), \nonumber
\end{eqnarray}
where the conditions of normalization (\ref{norm}) are reproduced.

Let us {\it postulate the extended definition of the vacuum}
\begin{equation}
|vac \rangle = |0_\tau \rangle \otimes |0_t \rangle \otimes |0_b \rangle ,
\label{vacuum}
\end{equation}
which implies the ${\mathbb Z}_3$ symmetry of the vacuum. Then, the couplings
introduced
in (\ref{generation}) are extended to three generations of fermions, whereas
the only generation of $\tau$, $t$ and $b$ is heavy, while two junior
generations are massless.

The vacuum definition (\ref{vacuum}) could be treated as the following
assumption:

\noindent
{\sf The number of generations equals the number of charged flavors in the
generation as well as the number of Higgs fields in the local phase.}

Thus, we postulate the ${\mathbb Z}_3$ symmetry of the vacuum as the
fundamental
dynamical principal of the theory. Moreover, we suggest that this symmetry of
the vacuum is exact, so that it is conserved under radiative corrections to the
Yukawa constants of fermions.

A realistic description of generations, i.e., a model with nonzero masses of
junior fermions is not the problem under the current consideration, and it is
beyond the scope of this work. Nevertheless, we add two notes.

First, a general structure of Yukawa interactions with the ${\mathbb Z}_3$
symmetry of
the vacuum has the form
\begin{eqnarray}
{\cal L}^\tau_{Y} & \sim & \bar \tau_R \tau_L \cdot (g_{1}^\tau
\varphi_1+g_{2}^\tau
\varphi_2+ g_{3}^\tau
\varphi_3) + {\rm h.c.},
\nonumber \\
{\cal L}^t_{Y} & \sim & \bar t_R t_L \cdot ( g_{1}^t \varphi_1+ g_{2}^t
\omega^2 \varphi_2+
g_{3}^t \omega \varphi_3)
+ {\rm h.c.}, \label{real}\\
{\cal L}^b_{Y} & \sim & \bar b_R b_L \cdot ( g_{1}^b \varphi_1+ g_{2}^b \omega
\varphi_2+
g_{3}^b \omega^2 \varphi_3)
+ {\rm h.c.}, \nonumber
\end{eqnarray}
where the constants $g_i$ can be restricted by the following conditions:
$g_{2,3}$ are real, while $g_1$ can be complex. So, the symmetric point $$
g_1=g_2=g_3=1,
$$
restores the hierarchy of single heavy generation and two massless generations.
The development of {\it ansatz} (\ref{real}) with the realistic values of
parameters consistent with the current data on the quark masses and mixing CKM
matrix of charged quark currents was given in Ref. \cite{generations}.

Second, the same form of mass matrix following from (\ref{real}) could result
in the leading order symmetry in the sector of neutrinos. Indeed, if we put
$$
g_1=e^{i\frac{2\pi}{3}},\;\;\;  |g_1|=g_2=g_3=1,
$$
then we get the completely degenerate neutrinos, while small deviations in the
$g_1$ phase and absolute values of $g_i$ will result in small differences of
neutrino masses squared as observed in the neutrino oscillations
\cite{neutrinos}.

\section{Conclusion}

In this work we have argued that there are three starting points for the
presented consideration of Higgs sector of electroweak theory. These
motivations are the following:

First, as well known, in the Standard Model the Yukawa coupling constant of $t$
quark obtained from the measurement of $t$ quark mass, is close to its value in
the infrared fixed point derived from the renormalization group equation. If
the Higgs sector is extended to three scalar fields separately coupled to
three heaviest charged fermions, i.e., the $t$, $b$ quarks and $\tau$ lepton,
with the same vacuum expectation values of Higgs fields, then the Yukawa
coupling constant of $t$ quark is {\bf exactly} posed to the infrared fixed
point. The challenge is whether this coincidence is accidental or not. We treat
this fact as the fundamental feature of dynamics determining the development of
masses. Moreover, the problem acquires an additional insight because the only
fermion generation is heavy, while two junior generations are approximately
massless. These features can be attributed by introducing the fundamental
${\mathbb Z}_3$
symmetry of the vacuum, so that this symmetry is conserved under the radiative
loop corrections responsible for the development of nonzero masses of junior
generations.

Second, the strong self-interaction regime in the Higgs sector of Standard
Model at large virtualities can be treated as the indication of nonlocality,
i.e., the compositeness of operators relevant to the electroweak symmetry
breaking. We have introduced a separation of virtuality regions: the local
Higgs phase in the range of $[0;\, \Lambda]$, the nonlocal strong
self-interaction regime in the range $[\Lambda;\, M]$, and the symmetric phase
above $M$. We have determined a form of composite operators and their
connection to the local phase by considering the second order of effective
action in the SM.

Third, the development of effective potential for the global sources of
composite operators from the point of symmetric phase $M\sim 10^{12}-10^{19}$
GeV is stopped in the infrared fixed point $\Lambda \approx 633$ GeV for the
Yukawa coupling constant of $t$ quark. If the dynamics of evolution is given by
the same electroweak group, then the large logarithm of $\ln M/\Lambda$ is
close to the value appearing in the calculation of GUT scale. So, since the
breaking of electroweak symmetry and the fermion mass generation involve the
composite operators with both left and right handed fermions, the gauge
interaction of ${\sf U}(1)$ group determines the evolution of parameters of the
effective potential in the strong self-interaction regime. Then the logarithm
of $\ln M/\Lambda$ in the coupling $g_1$ has the value depending on the set of
fundamental fields above the scale $\Lambda$. Anyway, $M$ should be close to
the
GUT scale, and this fact implies the solution of problem on the naturalness.

Then, following the above motivations, we have evaluated the basic parameters
of the model. We have calculated the effective potential for the sources of
composite operators, responsible for the breaking down the electroweak symmetry
and generation of masses for the gauge bosons and heaviest fermions. The
corresponding couplings serve as the matching values for the quadratic and
quartic constants in the potential of local Higgs fields as well as the Yukawa
interactions at the scale $\Lambda$, which is the ultraviolet cut-off for the
local theory and the low boundary of $[\Lambda;M]$-range for the effective
potential of sources coupled with the bi-local composite operators of quarks
and leptons. At $M$ the local gauge symmetry is restored, so that the effective
potential is exactly equal to zero.

Posing the matching of Yukawa constant for the $t$-quark to the infrared fixed
point at the scale $\Lambda$, related to the gauge boson masses, we have found
the null-potential value $M$ in the range of GUT park, which indicates the
solution of naturalness. The exploration of fixed points has resulted in the
following current masses of heaviest fermions: $m_t(m_t) = 165\pm 4$ GeV,
$m_b(m_b) = 4.18\pm 0.38$ GeV and $m_\tau(m_\tau) = 1.78\pm 0.27$ GeV. Two
degenerated neutral Higgs fields have the infrared fixed mass $m_H= 306\pm 5$
GeV, and the third scalar has the mass $m_{H_\tau}= 552\pm 9$ GeV. So, the
estimates do not contradict with the current constraints, coming from the
experimental data.

Some questions need for an additional consideration. To the moment, discussing
no possible ways to study, we focus on the directions requiring a progress.

\begin{itemize}
\item[1.]
What is a picture for the generation of Yukawa constants, responsible for the
masses of ``junior'' fermions?
\end{itemize}
As we have supposed in the paper, three sectors of Higgs fields are coupled to
the appropriate heavy fermions, so that we need speculations based on a
symmetry causing the junior generations to be massless to the leading order.
\begin{itemize}
\item[2.]
What are the constraints on the model parameters as follows from the current
data on the flavor changing neutral currents and precision measurements at LEP?
\end{itemize}
So, we expect that this point is not able to bring serious objections against
the model, since we do not involve any interactions distinct from the gauge
ones, composing the SM.
\begin{itemize}
\item[3.]
The most constructive question is a supersymmetric extension of mechanism under
consideration. Can SUSY provide new features or yield masses of super-partners?
\end{itemize}
To our opinion, the SUSY extension is more complicated, since there are many
different relations between the mixtures of various sparticles, which all are
expected to be essentially massive ($\tilde m\sim \Lambda$) in contrast to the
SM, wherein the junior generations are decoupled from the Higgs fields to the
leading order.
\begin{itemize}
\item[4.]
A simple application, we think, is an insertion of the model into the TeV-scale
Kaluza-Klein ideology, being under intensive progress now \cite{KK}.
\end{itemize}
So, $\varkappa(v)$ transforms its logarithmic behavior to the
power dependence on the scale. Then, the $M$ returns to a value
not far away from the matching point $\Lambda\sim 1$ TeV, as it
should be in the KK approach.

Thus, we have offered the model of electroweak symmetry breaking, which
provides a positive connection to the naturalness as well as needs some deeper
studies under progress.

This work is in part supported by the Russian Foundation for Basic Research,
grants 99-02-16558, 01-02-99315, 01-02-16585 and 00-15-96645, the Russian
Ministry on the education, grant E00-3.3-62.

\end{document}